\documentclass[prl,twocolumn,showpacs,superscriptaddress,nofootinbib,floatfix,10pt]{revtex4-1}
\usepackage{graphicx}
\usepackage{amsmath}
\usepackage{bm}
\usepackage{yhmath}
\usepackage{mathtools}
\usepackage{wasysym}
\usepackage[colorlinks,citecolor=blue,urlcolor=blue,linkcolor=blue]{hyperref}
\usepackage{subfigure}
\usepackage{color}
\usepackage{cases}
\usepackage{subfigure}
\usepackage{times}
\usepackage{dcolumn,booktabs,bm}
\usepackage{slashed}
\usepackage{amsfonts,amssymb,stmaryrd,latexsym,amsmath}
\usepackage{textcomp}
\usepackage{multirow}
\usepackage{cancel}
\usepackage{array}

\def\rmI{{\rm I}}
\def\rmII{{\rm I\!I}}

\newcommand{\etal}{\textit{et al}.}

\allowdisplaybreaks

\begin{document}

\title{$Z_{cs}(3985)^-$ as the $U$-spin partner of $Z_c(3900)^-$ and implication of other states\\ in the $\text{SU(3)}_F$ symmetry and heavy quark symmetry}

%\title{Strong evidence of $Z_{cs}(3985)^-$ as $U$-spin partner of charged $Z_c(3900)^-$}

\author{Lu Meng}
\affiliation{Center of High Energy Physics, Peking University,
    Beijing 100871, China} \affiliation{School of Physics and State Key
    Laboratory of Nuclear
    Physics and Technology, Peking University, Beijing 100871,
    China}\affiliation{Ruhr University Bochum, Faculty of Physics and Astronomy,
    Institute for Theoretical Physics II, D-44870 Bochum, Germany}

\author{Bo Wang}\email{bo-wang@pku.edu.cn}
\affiliation{Center of High Energy Physics, Peking University,
Beijing 100871, China} \affiliation{School of Physics and State Key
Laboratory of Nuclear
    Physics and Technology, Peking University, Beijing 100871,
    China}

\author{Shi-Lin Zhu}\email{zhusl@pku.edu.cn}
\affiliation{School of Physics and State Key Laboratory of Nuclear
Physics and Technology, Peking University, Beijing 100871,
China}\affiliation{Center of High Energy Physics, Peking University,
Beijing 100871, China}

\begin{abstract}
Recently, for the first time, the BESIII Collaboration reported the
strange hidden charm tetraquark states $Z_{cs}(3985)^-$ in the $K^+$
recoil-mass spectrum near the $D_{s}^{-}D^{*0}/ D_{s}^{*-}D^{0}$
mass thresholds in the processes of $e^{+}e^{-}\to
K^{+}(D_{s}^{-}D^{*0}+D_{s}^{*-}D^{0})$ at $\sqrt{s}=4.681$
GeV~\cite{1830518}. The significance was estimated to be 5.3
$\sigma$. We show that the newly observed $Z_{cs}(3985)^-$ state is
the $U$-spin partner of $Z_c(3900)^{-}$ as a resonance within
coupled-channel calculation in the $\text{SU(3)}_F$ symmetry and
heavy quark spin symmetry (HQSS). In the $\text{SU(3)}_F$ symmetry,
we introduce the $G_{U/V}$ parity to construct the flavor wave
functions of the $Z_{cs}$ states. In a unified framework, we
consider the $J/\psi\pi(K)$, $\bar{D}_{(s)}{D}^*/\bar{D}_{(s)}^*{D}$
coupled-channel effect with the contact interaction. With the masses
and widths of $Z_c(3900)$ and $Z_c(4020)$, we determine all the
unknown coupling constants. We obtain mass and width of
$Z_{cs}(3985)$ in good agreement with the experimental results,
which strongly supports the $Z_{cs}$ states as the $U/V$-spin
partner states of the charged $Z_c(3900)$. We also calculate the
ratio of the partial decay widths of $Z_{cs}(3985)$, which implies
that the $\bar{D}_sD^*/\bar{D}^*_s D$ decay modes are dominant. We
also predict the $Z_{cs}^\prime$ states with a mass around $4130$
MeV and width around $30$ MeV, which are the $U/V$-spin partner
states of the charged $Z_c(4020)$ and HQSS partner states of the
$Z_{cs}(3985)$. In the hidden bottom sector, we predict the strange
tetraquark states $Z_{bs}$ and $Z_{bs}^\prime$ with a mass around
10700 MeV and 10750 MeV, which are the $U/V$-spin partner states of
$Z_{b}(10610)^{\pm}$ and $Z_b(10650)^{\pm}$, respectively.
\end{abstract}

\maketitle

\thispagestyle{empty}

{\it Introduction.}---Very recently, the BESIII Collaboration
reported a novel structure $Z_{cs}(3985)^-$ in the $K^+$ recoil-mass
spectrum near the $D_{s}^{-}D^{*0}/ D_{s}^{*-}D^{0}$ mass thresholds
in the processes of $e^{+}e^{-}\to
K^{+}(D_{s}^{-}D^{*0}+D_{s}^{*-}D^{0})$ at $\sqrt{s}=4.681$
GeV~\cite{1830518}. The pole mass and width were determined with a
mass-dependent-width Breit-Wigner line shape,
\begin{eqnarray}
    M_{Z_{cs}}^{\text{pole}}&=&(3982.5_{-2.6}^{+1.8}\pm2.1) \text{ MeV},\nonumber\\
    \Gamma_{Z_{cs}}^{\text{pole}}&=&(12.8_{-4.4}^{+5.3}\pm3.0) \text{ MeV},
\end{eqnarray}
where the first and the second uncertainties are statistical and
systematic, respectively. The significance of the resonance
hypothesis is estimated to be 5.3 $\sigma$ over the pure
contributions from the conventional charmed mesons.

The minimum quark constituents of $Z_{cs}(3985)$ are ($c\bar{c}s
\bar{n}$), where $n$ represents the $u/d$ quark. While the number of
exotic states is rapidly growing (see
Refs.~\cite{Brambilla:2019esw,Liu:2019zoy,Guo:2017jvc,Olsen:2017bmm,Chen:2016qju,Esposito:2016noz}
for recent reviews), $Z_{cs}(3985)$ is still a very unusual state by
current standards. Most XYZ states are isospin singlet, in which the
numbers of constituent quark are not fixed. The unquenched quark
dynamics~\cite{Barnes:2007xu,Eichten:2004uh} would mix the two quark
components with four quark components. However, the charged
$Z_c$/$Z_b$ states~\cite{Belle:2011aa,Ablikim:2013mio,Liu:2013dau},
the $P_c$ states~\cite{Aaij:2019vzc,Aaij:2015tga} and the newly
observed $Z_{cs}(3985)$ are multiquark states without much doubt. In
addition, $Z_{cs}(3985)$ might be the rare hidden charm exotic
candidate with strange number. Another candidate is the $P_{cs}$
states reported recently by LHCb Collaboration~\cite{Wanglhcb:2020}.

The $Z_c(3900)$ and $Z_c(4020)$ states are above the threshold of
$\bar{D}^*D/\bar{D}D^*$ and $\bar{D}^*D^*$ by several MeVs,
respectively. $Z_c(4020)$ states are likely the heavy quark spin
symmetry (HQSS) partners of the $Z_c(3900)$ states. The theoretical
interpretations of $Z_c(3900)$ and $Z_c(4020)$ range from the
threshold effect~\cite{Swanson:2014tra}, to compact tetraquark
states~\cite{Dias:2013xfa,Agaev:2017tzv}, or hadronic
molecules~\cite{Guo:2013sya,He:2013nwa}. The threshold effect
picture of $Z_c$ states was challenged by Guo
\etal~\cite{Guo:2014iya} and JPAC
Collaboration~\cite{Pilloni:2016obd}. In the tetraquark scheme, it
is hard to understand their proximity to the di-meson thresholds. In
molecular scenario, the one-pion-exchange interaction for $I=1$
$\bar{D}^{(*)}D^{(*)}$ systems is mainly repulsive. Therefore,
theorists resorted to the coupled-channel calculation to interpret
$Z_c$ states as the molecular-type resonances, virtual states or
bound
states~\cite{Ortega:2018cnm,Chen:2019iux,Ikeda:2016zwx,Lee:2014uta,Prelovsek:2014swa,Aceti:2014uea,He:2015mja,He:2017lhy,Albaladejo:2015lob,Hanhart:2015cua,Guo:2016bjq,Wang:2018jlv,Wang:2020dko}.

In the history, the successful prediction of $ \Omega^{-}$ taught us
the importance of $\text{SU(3)}_F$ symmetry~\cite{Barnes:1964pd} in
hadron spectroscopy. In the exotic hadron sector, within a dominant
short-range interaction from $\text{SU(3)}_F$ symmetry, we predicted
the strange hidden charm pentaquark state as the $\Xi_c\bar{D}^*$
bound state with a mass $4456.9$ MeV~\cite{Wang:2019nvm}. Recently,
our prediction was supported by the observation of $P_{cs}(4459)^0$
by LHCb collaboration~\cite{Wanglhcb:2020}. The $Z_c(3900)$ and
$Z_{cs}(3985)$ states are in the proximity of the threshold of
$\bar{D}D^*/\bar{D}^*D$ and $\bar{D}_s D^*/\bar{D}^*_{s}D$,
respectively. It is natural to conjecture that they belong to the
same $\text{SU(3)}_F$ multiplet. Therefore, it is crucial to
investigate $Z_c(3900)/Z_c(4020)$ and the newly observed
$Z_{cs}(3985)$ state in a unified framework with $\text{SU(3)}_F$
symmetry and HQSS. In this Letter, we first make use of the HQSS to
interpret the $Z_c(3900)$ and $Z_c(4020)$ as resonances in
$J/\psi\pi$, $\bar{D}D^*/\bar{D}^*D$ and $\bar{D}^*D^*$
coupled-channel calculation. Then, within the $\text{SU(3)}_F$
symmetry, we extend the calculation to the strange channels without
unknown parameters. We aim at obtaining the mass and width of
$Z_{cs}$ state and its HQSS partner state. Meanwhile, the $Z_{cs}$
was firstly observed in $\bar{D}_sD^*/\bar{D}^*D_s$ channel rather
than in the hidden channels like $J/\psi K$. Another question
addressed in this Letter is to determine the dominant decay channels
of $Z_{cs}(3985)$ and its possible HQSS partners.

{\it $U/V$-spin partners of $Z_c(3900)^{\pm}$ and
$Z_c(4020)^{\pm}$.}---The quantum numbers of $Z_c(3900)$ are
$I^G(J^{PC})=1^+(1^{+-})$ ($C$ parity only for the neutral states
here and below)~\cite{Zyla:2020zbs}. For the S-wave
$\bar{D}^*D/\bar{D}D^*$ channel, we could construct two orthogonal
basis vectors,
 \begin{eqnarray}
    \frac{1}{\sqrt{2}}\left(|\bar{D}D^{*}\rangle+\eta|\bar{D}^{*}D\rangle\right),\label{eq:dd}
 \end{eqnarray}
where $J^{PC}=1^{+\pm}$ for $\eta=\mp1$. We omit the isospin
information in Eq.~\eqref{eq:dd}. For the $I=1$ channels, the
$G$-parity (eigenvalue of $\hat{G}=\hat{C}e^{i\hat{I}_2\pi}$) is
$\eta$. Thus, the $Z_c(3900)$ states correspond to the isovector
channel with $\eta=+1$ in Eq.~\eqref{eq:dd}. The quantum numbers of
$Z_c(4020)$ are $I^G(J^{PC})=1^+(?^{?-})$~\cite{Zyla:2020zbs}. As
the HQSS partner states of $Z_c(3900)$, $Z_c(4020)$ states will
couple with the S-wave $\bar{D}^*D^*$ isovector channel, which
implies its possible $J^P$ could be $0^+,1^+,2^+$. We will assume
the $J^P$ of $Z_c(4020)$ is $1^+$ and the reason will be given
later.

\begin{figure}[!htp]
    \centering
    \includegraphics[width=0.45\textwidth]{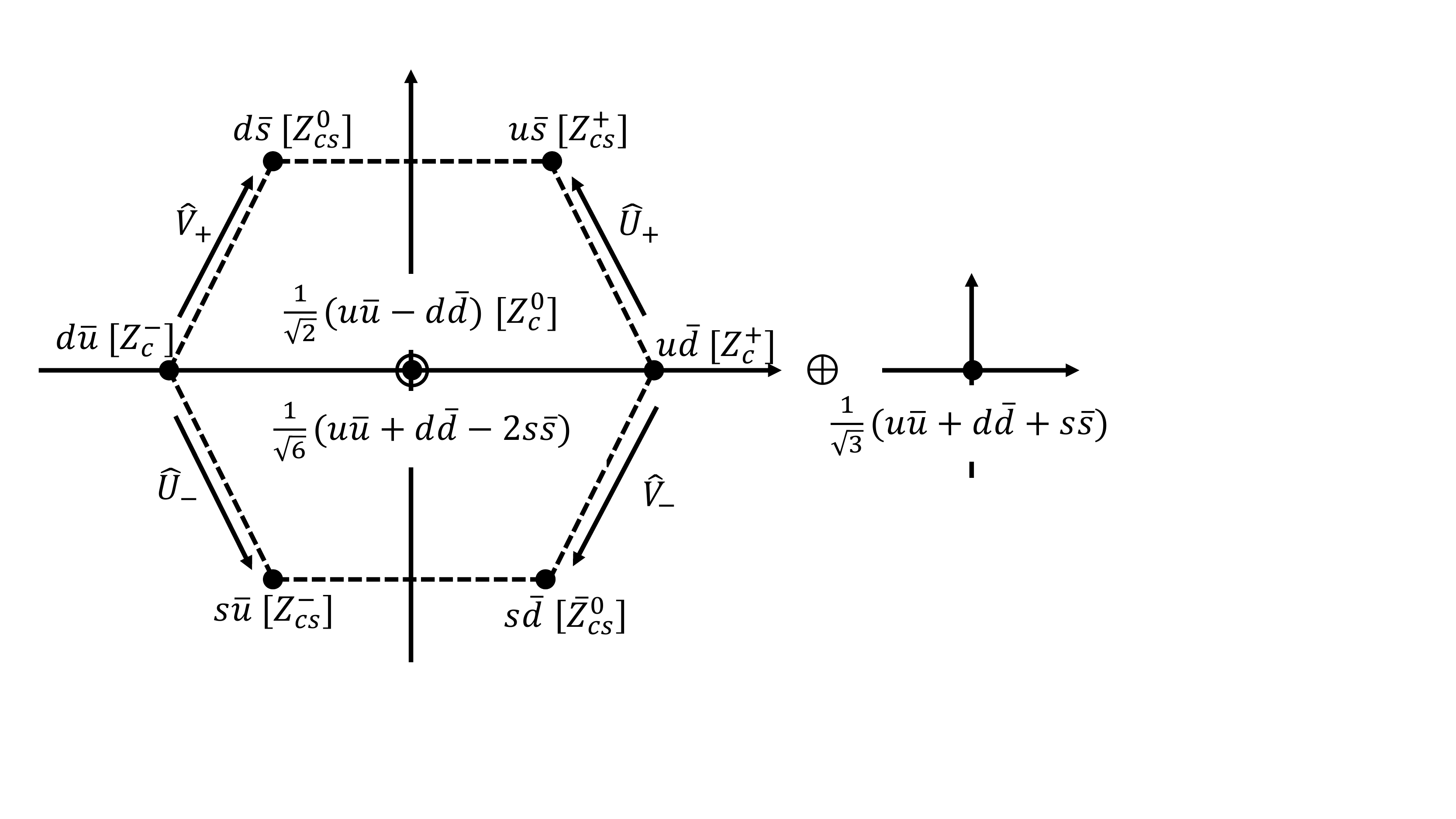}
    \caption{The multiplet structure of the $\bar{D}^{(*)}_{(s)}{D}^{(*)}_{(s)}$ di-meson systems in the $\text{SU(3)}_F$ symmetry. We omit the heavy quarks in flavor wave functions for conciseness. }\label{fig:weight}
\end{figure}

We assume $Z_{cs}(3985)$ state is the strange partner of $Z_c(3900)$
in the $\text{SU(3)}_F$ symmetry. To be specific, the $Z_{cs}$
states are related to the $Z_c$ states with the rotation in
$U/V$-spin space as shown in Fig.~\ref{fig:weight},
\begin{equation}
    Z_{c}^{-} \xleftrightarrow{U} Z_{cs}^{-},\quad Z_{c}^{+} \xleftrightarrow{V} \bar{Z}_{cs}^{0}.
\end{equation}
$U$-spin and $V$-spin are the SU(2) subgroups of the SU(3) group
just like the isospin subgroup. The SU(2) doublets for these
subgroup are
\begin{equation}
    u,d\:(I);\quad d,s\:(U);\quad u,s\:(V).
\end{equation}

The thresholds of $D_s^-D^{*0}$ (3975 MeV) and $D_s^{*-}D^0$ (3977
MeV) are very close. In the heavy quark limit, $\bar{D}_sD^*$ and
$\bar{D}_s^*D$ are degenerate. We construct the basis of the
di-meson channel $\bar{D}_sD^*/\bar{D}_s^*D$ like Eq.~\eqref{eq:dd},
\begin{eqnarray}
    |G_{V}=\eta\rangle&=\frac{1}{\sqrt{2}}\left(|{D}_{s}^{-}{D}^{*0}\rangle+\eta|{D}_{s}^{*-}{D}^{0}\rangle\right),\nonumber\\
    |G_{U}=\eta\rangle&=\frac{1}{\sqrt{2}}\left(|D_{s}^{-}D^{*+}\rangle+\eta|D_{s}^{*-}D^{+}\rangle\right),\label{eq:dsd}
\end{eqnarray}
where $\hat{G}_U$ and $\hat{G}_V$ transformations are defined like
$\hat{G}$,
\begin{equation}
    \hat{G}_U=\hat{C}e^{i\hat{U}_2\pi},\quad \hat{G}_V=\hat{C}e^{i\hat{V}_2\pi}.
\end{equation}
The di-meson channels with $\eta=+1$ correspond to $Z_{cs}^{-}$ and
$\bar{Z}_{cs}^0$ with $G_{U/V}=+1$. Similarly, we construct the
$\bar{D}_s^*D^*$ di-meson channels with $J^P=1^+$ and $G_{U/V}=+1$,
which are the HQSS partner channels of Eq.~\eqref{eq:dsd} with
$\eta=+1$. These channels correspond to the HQSS partner states of
$Z_{cs}$ and $U/V$-spin partner states of $Z_c(4020)$, which are
named as $Z_{cs}^{\prime}$ here and below.

The dynamics of the $\bar{D}^{(*)}_{(s)}{D}^{(*)}_{(s)}$ di-meson
systems are constrained by both $\text{SU(3)}_F$ symmetry and HQSS.
In the heavy quark limit, the $c$ and $\bar{c}$ are the spectators
in the di-meson systems. For the $S$-wave channel, $\mathbb{I}_s$
and $\bm{l}_1 \cdot \bm{l}_2$ are the only interaction operators in
spin space, where $\mathbb{I}_s$ is the unit operator in spin space
and $\bm{l}_i$ is the light spin operator of the heavy meson. In the
$\text{SU(3)}_F$ symmetry, the interaction operators in flavor space
between light degrees of freedom are $\mathbb{I}_F$ and
$\mathbb{C}_2$, where $\mathbb{I}_F$ is the unit operator and
$\mathbb{C}_2=-\sum_{i=1}^8 \lambda_F^i\lambda_F^{*i}$ is the
Casimir operator. Therefore, the general interaction for
$\bar{D}^{(*)}_{(s)}{D}^{(*)}_{(s)}$ could be parameterized as,
\begin{equation}
V_{q\bar{q}}=c_{1}+c_{2}\bm{l}_{1}\cdot\bm{l}_{2}+c_{3}\mathbb{C}_{2}+c_{4}(\bm{l}_{1}\cdot\bm{l}_{2})\mathbb{C}_{2}.\label{eq:vqq}
\end{equation}

In the $\text{SU(3)}_F$ symmetry, the
$\bar{D}^{(*)}_{(s)}{D}^{(*)}_{(s)}$ systems could be classified by
$3_F\otimes \bar{3}_F \to  8_F\oplus 1_F$ as shown in
Fig.~\ref{fig:weight}. The matrix elements of the Casimir operator
$\mathbb{C}_2$ read:
\begin{equation}
\langle \mathbb{C}_2 \rangle_{8_F}={2\over 3},\quad \langle
\mathbb{C}_2 \rangle_{1_F}=-{16\over 3}.
\end{equation}
Thus, in the $\text{SU(3)}_F$ symmetry, the $\bar{D}^{(*)}D^{(*)}$
interactions for the $Z_c(3900)$ and $Z_c(4020)$ are the same as
those of $\bar{D}^{(*)}_sD^{(*)}$ concerning with the $Z_{cs}$
states.

In the spin space, we write the spin wave function of
$\bar{D}^{(*)}_{(s)}{D}^{(*)}_{(s)}$ as $|l_1 h_1(j_1)l_2
h_2(j_2)JM\rangle$, where $l_i$, $h_i$ and $j_i$ are the light spin,
heavy spin and total spin of the heavy meson with label $i$. $J$ and
$M$ are the total spin and its third component of the di-meson
system. One can use the $9j$ symbols to relate the above spin wave
function to state $|l_1l_2(L)h_1h_2(H)JM\rangle $, where $L$ and $H$
are total light spin and total heavy spin, respectively. The matrix
elements of $\bm{l}_1\cdot\bm{l}_2$ can be calculated,
\begin{eqnarray}
&&\langle\bm{l}_{1}\cdot\bm{l}_{2}\rangle_{\{\mathtt{PP},\mathtt{VV}\}}^{0^{+}}=\left[\begin{array}{cc}
0 & \frac{\sqrt{3}}{4}\\
\frac{\sqrt{3}}{4} & -\frac{1}{2}
\end{array}\right],\label{eq:ll0}\\
&&\langle\bm{l}_{1}\cdot\bm{l}_{2}\rangle_{\{\mathtt{PV}\eta=+1,\mathtt{VV}\}}^{1^{+}}=\left[\begin{array}{cc}
-\frac{1}{4} & -\frac{1}{2}\\
-\frac{1}{2} & -\frac{1}{4}
\end{array}\right],\label{eq:ll}\\
&&\langle\bm{l}_{1}\cdot\bm{l}_{2}\rangle_{\{\mathtt{PV}\eta=-1\}}^{1^{+}}=\frac{1}{4},\quad
\langle\bm{l}_{1}\cdot\bm{l}_{2}\rangle_{\{\mathtt{VV}\}}^{2^{+}}=\frac{1}{4},\label{eq:ll2}
\end{eqnarray}
where $\mathtt{P}$ and $\mathtt{V}$ denote the pseudoscalar and
vector heavy mesons, receptively. We use the superscript to denote
the $J^P$ of the di-meson channel. For the $\mathtt{PV}$ channels,
we write the $\eta$ of Eqs.~\eqref{eq:dd} and \eqref{eq:dsd}
explicitly. Eqs.~\eqref{eq:ll0}-\eqref{eq:ll2} are the results in
the heavy quark limit. We can see that among the $\mathtt{VV}$
channels, the $J^P=1^+$ one has the same matrix elements with the
$\mathtt{PV}\eta=+1$ channel, which couples with the $Z_c(3900)$.
Thus, it is reasonable to assume the $J^P=1^+$ for $Z_c(4020)$. The
matrix elements in spin space of channels corresponding to
$Z_c(3900)/Z_{cs}(3985)$ and $Z_c(4020)/Z_{cs}^{\prime}$ are equal.

The di-meson channels corresponding to $Z_c(3900)$, $Z_c(4020)$,
$Z_{cs}(3985)$ and $Z_{cs}^{\prime}$ have the same interaction,
which is the result of the $\text{SU(3)}_F$ symmetry and HQSS. We
embed these symmetries in Eq.~\eqref{eq:vqq}. One could adopt other
equivalent approaches like superfield method in
Ref.~\cite{Nieves:2012tt}.

{\it Coupled-channel calculation.}---We list the channels considered
in Table~\ref{tab:channel}. Apart from the
$\bar{D}^{(*)}_{(s)}{D}^{(*)}_{(s)}$ channels, we also include the
$J/\psi \pi$ channel for the $Z_c(3900)/Z_c(4020)$ systems and
$J/\psi K$ channel for the $Z_{cs}(3985)/Z_{cs}^{\prime}$ systems.
The coupled-channel $T$-matrix can be obtained by solving the
Lippmann-Schwinger equations (LSEs):
\begin{equation}
T_{ij}=V_{ij}+\sum_{k}V_{ik}G_{k}T_{kj},
\end{equation}
The loop function $G_i$ reads~\cite{Oller:1997ti},
\begin{equation}
G_i(E)=\int_0^{\Lambda_i}{l^2 dl \over (2\pi)^2} {w_{i1}+w_{i2}
\over w_1w_2[E^2-(w_{i1}+w_{i2})^2+i\epsilon]},\label{eq:green}
\end{equation}
where $w_{ia}=(\bm l^2+m_{ia}^2)^{1/2}$ and $m_{ia}$ is the mass of
the $a$-th particle in the channel $i$. We take a hard cutoff
$\Lambda_i$ to regulate the integral. We vary the cutoff parameters
$\Lambda_2=\Lambda_3=0.5-1.0$ GeV but keep the same $\Lambda_{2}$
and $\Lambda_3$ to avoid the unintentional HQSS breaking effect. For
the definiteness, we fix $\Lambda_1=1.5$ GeV. For the $Z_{cs}$ and
$Z_c$ systems, we choose the same cutoff parameters to keep the
$\text{SU(3)}_F$ symmetry.

\begin{table}
    \caption{Channels considered in the coupled-channel calculation.}\label{tab:channel}
        \setlength{\tabcolsep}{2.8mm}
    \begin{tabular}{cccc}
        \hline \hline
        Channel & 1 & 2 & 3\tabularnewline
        \hline
        $Z_{c}/Z_{c}^{\prime}$ & $J/\psi\pi$ & $\frac{1}{\sqrt{2}}\left(|\bar{D}D^{*}\rangle+|\bar{D}^{*}D\rangle\right)$ & $\bar{D}^{*}D^{*}$\tabularnewline
        $Z_{cs}/Z_{cs}^{\prime}$ & $J/\psi K$ & $\frac{1}{\sqrt{2}}\left(|\bar{D}_{s}{D}^{*}\rangle+|\bar{D}_{s}^{*}{D}\rangle\right)$ & $\bar{D}_{s}^{*}{D}^{*}$\tabularnewline
        \hline \hline
    \end{tabular}
\end{table}

Following the pionless effective field
theory~\cite{Epelbaum:2008ga,Machleidt:2011zz,Epelbaum:2008ga}, we
only introduce the contact interaction. For the off-diagonal
potential $V_{23}=V_{32}$, we take the leading order contact
interaction as a constant $v_{23}$. For the diagonal potential, we
have $V_{22}=V_{33}$ from Eq.~\eqref{eq:ll}. In order to obtain the
resonances above thresholds, we introduce the next-to-leading order
contact interaction for the elastic
potential~\cite{Albaladejo:2015lob},
\begin{equation}
V_{22}=V_{33}=C_{d}+{C_{d}^{\prime}\over 2}(\bm{p}^2+\bm{p}^{\prime
2}),
\end{equation}
where $\bm{p}$ and $\bm{p}^{\prime}$ are the initial and final
momenta in the center-of-mass system (c.m.s). The general terms at
the next-to-leading order are $(\bm{p}+\bm{p}^{\prime})^2$ and
$(\bm{p}-\bm{p}^{\prime})^2$, while the $\bm{p}\cdot\bm{p^{\prime}}$
term vanishes after partial wave expansion for the $S$-wave. When
the particles are on-shell, the magnitude $p_{i}=|\bm{p}_i|$ of
channel $i$ in c.m.s is
\begin{equation}
p_{i}(E)=\frac{\sqrt{[E^{2}-(m_{i1}+m_{i2})^{2}][E^{2}-(m_{i1}-m_{i2})^{2}]}}{2E}.
\end{equation}

The elastic interaction for $J/\psi \pi $ or $J/\psi K$ is purely
gluonic vander Waals force, which is known to be
tiny~\cite{Yokokawa:2006td,Liu:2008rza,Liu:2012dv}. We neglect the
diagonal interaction in the first channel, $V_{11}=0$. The processes
$\bar{D}^{(*)}D^{(*)}\to J/\psi \pi$ and $\bar{D}^{(*)}_sD^{(*)}\to
J/\psi K$ are related by the $U/V$-spin transformation. Thus the
$V_{1i}$ for strange systems and non-strange systems are the same.
In the heavy quark limit, the channel 2 and channel 3 have the same
spatial wave function and flavor wave function, thus we focus on the
spin wave function. The ratio $V_{12}/V_{13}$ could be estimated by
ratio of spin wave function overlaps,
\begin{equation}
\frac{V_{12}}{V_{13}}=\frac{\langle J/\psi \pi
|\mathtt{PP}\eta=+1,1^{+}\rangle_{\text{spin}}}{\langle J/\psi \pi
|\mathtt{VV},1^{+}\rangle_{\text{spin}}}=1.\label{eq:v12v13}
\end{equation}
With Eq.~\eqref{eq:v12v13}, we can parameterize the $V_{12}$ and
$V_{13}$ with one single coupling constant $v_{12}$.

With the HQSS and $\text{SU(3)}_F$ symmetries, the $V_{ij}$ reads
\begin{equation}
V_{ij}=\left[\begin{array}{ccc}
0 & v_{12} & v_{12}\\
v_{12} & C_{d}+\frac{C_{d}^{\prime}}{2}(\bm{p}^{2}+\bm{p}^{\prime2}) & v_{23}\\
v_{12} & v_{23} &
C_{d}+\frac{C_{d}^{\prime}}{2}(\bm{p}^{2}+\bm{p}^{\prime2})
\end{array}\right].
\end{equation}
We have four unknown coupling constants, $v_{12}$, $v_{23}$, $C_d$
and $C_d^{\prime}$. We shall solve the LSEs and fit the masses and
widths of $Z_c(3900)$ and $Z_c(4020)$ to determine the four coupling
constants. The resonances are located in the unphysical Riemann
sheet which is accessed by analytical
continuation~\cite{Nieves:2001wt,Oller:1997ti}. We replace $G_i(E)$
with
\begin{equation}
G_i^{\rmII}(E)=G_i^\rmI(E)+i {p_i(E) \over 4\pi E},
\end{equation}
where $G_i^{\rmI}$ is the loop function in Eq.~\eqref{eq:green}.

Since the widths of these resonances are narrow, $\Gamma \ll M$, we
could estimate the partial decay widths with the Breit-Wigner
parameterization~\cite{Oller:1997ti}. The $T_{ij}(E)$ matrix reads,
\begin{equation}
T_{ij}(E)={1\over 2 M_R} {g_i(E)g_j(E) \over E-M_R+i {\Gamma_R \over
2}},
\end{equation}
where $M_R$ and $\Gamma_R$ are the mass and width of the resonance,
respectively. $g_i$ is the coupling vertex of the resonance and
particles in channel $i$. The partial decay width $\Gamma_i$ reads,
\begin{equation}
\Gamma_{i}=\int\frac{1}{2M_{R}}|{\cal M}_{R\to
i}|^{2}2\pi\delta(M_{R}-E_{i1}-E_{i2})\frac{d^{3}\bm{p}_{i1}}{(2\pi)^{3}2E_{i1}2E_{i2}},
\end{equation}
where $\mathcal{M}_{R \to i}=g_i$. We make the substitution in the
narrow-width approximation,
\begin{equation}
2\pi\delta(M_{R}-E_{i1}-E_{i2})\to2\text{Im}\frac{1}{M_{R}-E_{i1}-E_{iR}-i\frac{\Gamma_{R}}{2}}.
\end{equation}
We change the integral variable to $E$ and the partial wave decay
width becomes:
\begin{eqnarray}
\Gamma_{i}
%&=&\frac{1}{16\pi^{2}}\int_{m_{i1}+m_{i2}}^{\infty}dE\frac{p_{i1}(E)}{E^{2}}|g_{i}(E)|^{2}\frac{\Gamma_{R}}{(M_{R}-E)^{2}+\frac{\Gamma_{R}^{2}}{4}} \nonumber\\
&=&-\frac{1}{16\pi^{2}}\int_{m_{i1}+m_{i2}}^{\infty}dE\frac{p_{i1}(E)}{E^{2}}4M_{R}\text{Im}T_{ii}(E).
\end{eqnarray}
In practical calculation, we integrate in $E$ around 2 widths up and
down the pole mass (if allowed by the lower limit ) to obtain
$\Gamma_{i}$, since we find $\sum_i \Gamma_i \approx\Gamma_{R}$ in
this integration range.

\begin{figure}[!htp]
    \centering
    \includegraphics[width=0.45\textwidth]{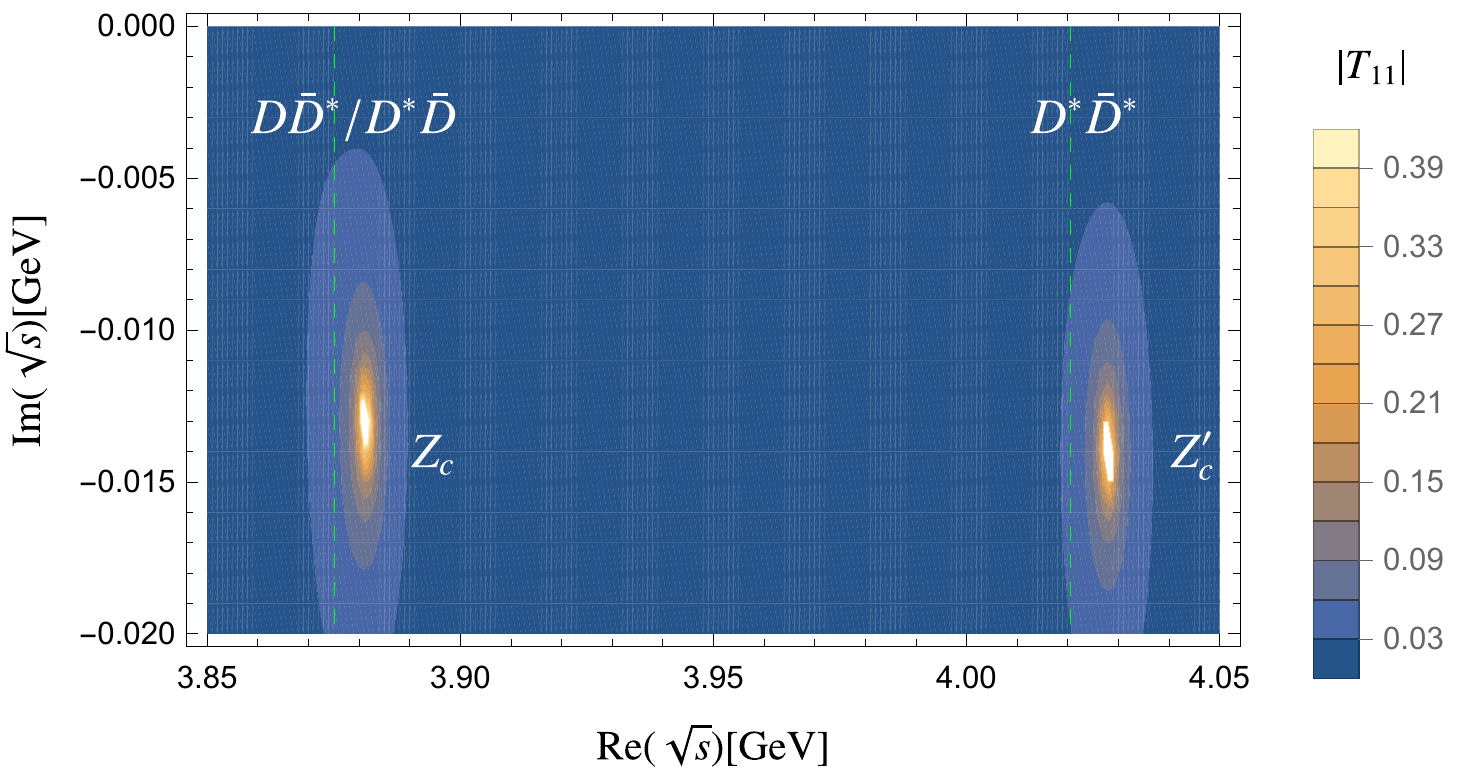}
    \includegraphics[width=0.45\textwidth]{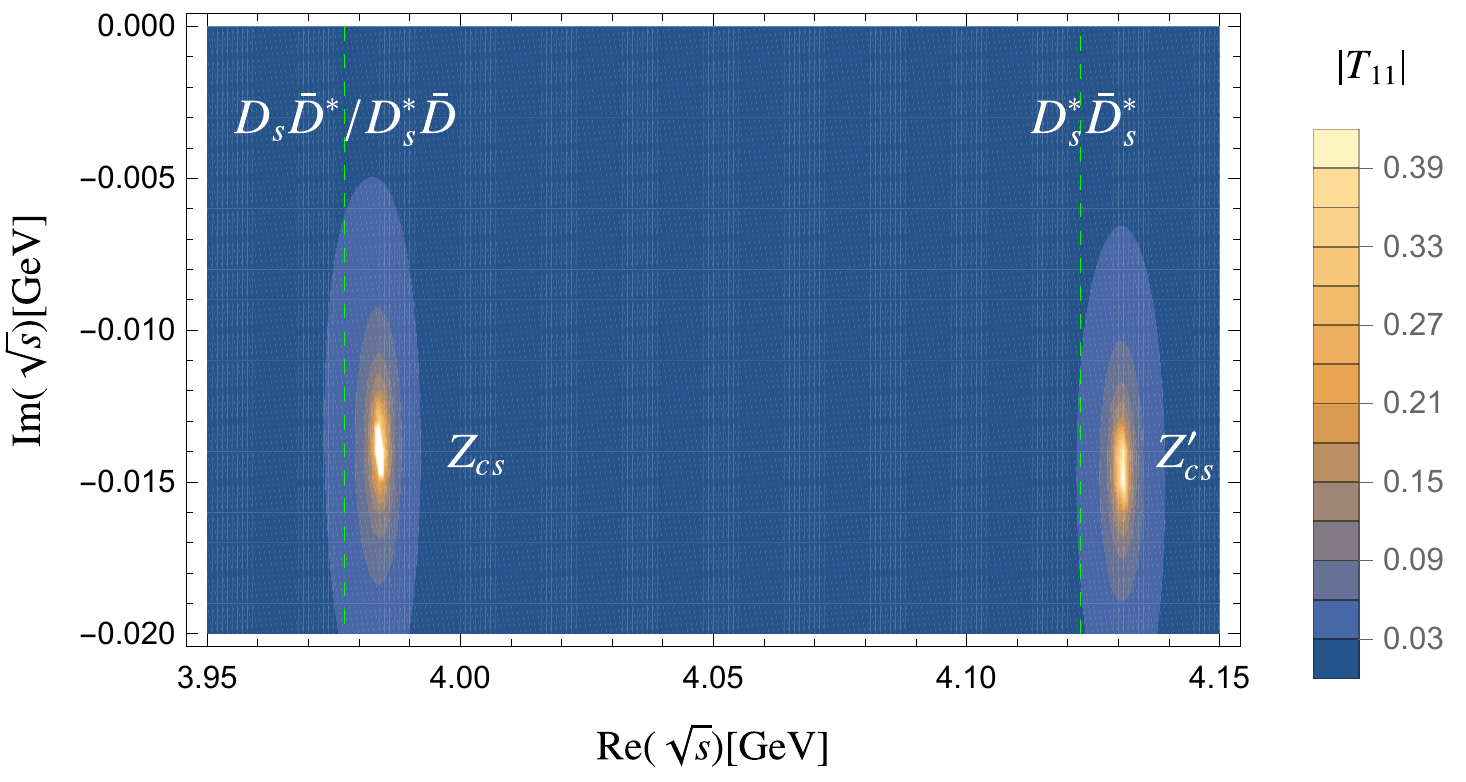}
    \caption{$T_{11}$-matrix in the unphysical sheet with $\Lambda_{2/3}=1.0$ GeV and pole positions . The upper plot and the lower plot are for the non-strange and strange states, respectively. In each plot, the dashed lines represent the thresholds of $\bar{D}_{(s)}{D}^*/\bar{D}_{(s)}^*D$ and $\bar{D}^*_{(s)}{D}^*.$ }\label{fig:zcpole}
\end{figure}

{\it Numerical results and Discussions.}---We choose the recent
results of the charged $Z_c(3900)$ and $Z_c(4020)$ in
Refs.~\cite{Ablikim:2013emm,Ablikim:2015swa} as input. We could
either choose the averaged results in Ref.~\cite{Zyla:2020zbs},
which would give the similar final results. We determine the
coupling constants in either $\Lambda_{2/3}=1.0$ GeV or $0.5$ GeV
and then calculate the masses and widths of $Z_{cs}$ and
$Z_{cs}^{\prime}$. We present $T_{11}$-matrix in the unphysical
sheet with $\Lambda_{2/3}=1.0$ GeV in Fig.~\ref{fig:zcpole}. We can
see two poles corresponding to $Z_c(3900)$ and $Z_c(4020)$, which
are barely above the thresholds of $\bar{D}D^*$ and $\bar{D}^*D^*$
by several MeVs, respectively. The positions of poles are
$M-i\Gamma/2$, where $M$ and $\Gamma$ are the mass and width of
resonances.

\begin{table*}
    \caption{Numerical results for masses, widths and partial widths. We use ``$\dagger $'' to label input. The ratios $\Gamma_3/\Gamma_2$ are estimated with central values of coupling constants. The lower limit of ratios $\Gamma_i/\Gamma_1$ are estimated with upper limits of $v_{12}$. $M$ and $\Gamma$ are in unites of MeV and $\Lambda_i$ are in unites of GeV.}\label{tab:num}
    \setlength{\tabcolsep}{0.5mm}
    \begin{tabular}{c|cccc}
        \hline \hline
        $(M,\Gamma)$ & $Z_{c}(3900)$ & $Z_{c}(4020)$ & $Z_{cs}(3985)$ & $Z_{cs}^{\prime}$\tabularnewline
        \hline
        Exp.~\cite{1830518,Ablikim:2013emm,Ablikim:2015swa} & $(3881.7\pm2.3,26.6\pm2.9)^\dagger$ & $(4026.3\pm4.5,24.8\pm9.5)^\dagger $ & $(3982.5_{-2.6}^{+1.8}\pm2.1, 12.8_{-4.4}^{+5.3}\pm3.0) $ & \tabularnewline
        \hline
        $\Lambda_{2/3}=1.0$     & $(3881.3\pm3.3,26.3\pm6.1)$ & $(4028.0\pm2.6,28.0\pm6.5)$ & $(3984.2\pm3.3,27.6\pm7.3)$ & $(4130.7\pm2.5,29.1\pm6.4)$\tabularnewline
        & ${\Gamma_{2}\over \Gamma_{1}}\apprge 13.7$ & ${\Gamma_{3}\over \Gamma_{2}}\approx0.51,\:{\Gamma_{3}\over\Gamma_{1}}\apprge 12.1$ & ${\Gamma_{2}\over \Gamma_{1}}\apprge16.1$ & ${\Gamma_{3}\over \Gamma_{2}}\approx0.48,\:{\Gamma_{3}\over \Gamma_{1}}\apprge13.7$\tabularnewline
    %   & \multicolumn{4}{c}{$C_{d}=-119.0\pm5.2,\quad C_{d}^{\prime}=-1215.5\pm137.3\text{GeV}^{-2},\quad v_{12}=0.8\pm24.9,\quad v_{23}=0.0038\pm21.8$}\tabularnewline
        \hline
    $\Lambda_{2/3}=0.5$     & $(3881.5\pm3.5,26.4\pm5.8)$ & $(4027.3\pm3.3,27.0\pm6.7)$ & $(3983.7\pm4.1,26.7\pm5.8)$ & $(4129.4\pm3.3,27.3\pm9.2)$\tabularnewline
        & ${\Gamma_{2}\over \Gamma_{1}}\apprge11.2$ & ${\Gamma_{3}\over\Gamma_{2}}\approx2.5,\:{\Gamma_{3}\over\Gamma_{1}}\apprge11.0$ & $ {\Gamma_{2}\over\Gamma_{1}}\apprge12.8$ & ${\Gamma_{3}\over \Gamma_{2}}\approx2.3,\:{\Gamma_{3}\over\Gamma_{1}}\apprge11.6$\tabularnewline
%       & \multicolumn{4}{c}{$C_{d}=-155.7\pm14.0,\quad C_{d}^{\prime}=-3904.0\pm388.1\text{GeV}^{-2},\quad v_{12}=26.4\pm30.0,\quad v_{23}=1.1\pm125.2$}\tabularnewline
        \hline \hline
    \end{tabular}
\end{table*}

We give the numerical results in Table~\ref{tab:num}. One can see
that we could reproduce the mass and width of the newly observed
$Z_{cs}(3985)$ state with $Z_c(3900)$ and $Z_c(4020)$ as input. Our
results are in good agreement with the experimental results, which
strongly supports that isospin doublet $Z_{cs}(3985)$ states are the
$U/V$-spin partner of the charged $Z_c(3900)$ as resonances.
Meanwhile, we predict a new resonance above the threshold of the
$\bar{D}_s^*D^*$ by $8$ MeV with a width about 30 MeV, which is the
HQSS partner of the $Z_{cs}(3985)$ and $U/V$-spin partner of
$Z_c(4020)$.

In this calculation, we use the decay modes $J/\psi \pi(K)$,
$\bar{D}_{(s)}D^*/\bar{D}_{(s)}^*{D}$, $\bar{D}_{(s)}^*D^*$ to
saturate the total widths, which would bring some uncertainties.
These uncertainties would be compensated in ratios of partial decays
widths. The coupling constants $v_{12}$ is very small and thus the
resonances are dominated by the
$\bar{D}_{(s)}D^*/\bar{D}_{(s)}^*{D}$ and $\bar{D}^*_{(s)}{D}^*$
components. With the central value of $v_{12}$, the partial wave
decay widths of $\Gamma_1$ are very small, thus we take the upper
limit of $v_{12}$ to give the lower limit of
$\Gamma_{i}/\Gamma_{1}$. From Table~\ref{tab:num}, we can see that
the decay process to $J/\psi \pi(K)$ is suppressed by at least one
order compared with the open charmed final state decays. The
dominant $\bar{D}_{(s)}D^*/\bar{D}_{(s)}^*{D}$ decay modes lead to
the observation in these channels in experiment~\cite{1830518}.
Meanwhile, as shown in Eq.~\eqref{eq:dsd}, the $Z_{cs}(3985)$ states
with $G_{V/U}=1$ have the same components of $\bar{D}_sD^*$ and
$\bar{D}^*_s{D}$, which is constrained by the $\text{SU(3)}_F$
symmetry.

We have assigned seven states $Z_c^{0/\pm}$, $Z_{cs}^{0/+}$,
$\bar{Z}_{cs}^{0/-}$ into the SU(3) octet in Fig.~\ref{fig:weight}.
The left isospin singlet in $8_F$ representation might mix with the
isospin singlet in $1_F$ like the $\phi$ and $\omega$ mesons. The
$\mathbb{C}_2$ matrix elements read
$$\langle  \mathbb{C}_2\rangle_{s\bar{s}}=-{4\over 3},\quad  \langle\mathbb{C}_2\rangle_{(u\bar{u}+d\bar{d})/\sqrt{2}}=-{10 \over 3}.$$
Both matrix elements have different signs from those of the octet in
ideal SU(3) symmetry. Therefore, the mixture would make the eighth
state disappear. In the compact tetraquark scheme, the existence of
the tetraquark states do not depend on the flavor. Searching for the
eighth state would help to distinguish the compact tetraquark states
from the di-meson states.

We further extend the calculation to the hidden bottom sector with
heavy quark flavor symmetry (HQFS). We use the coupling constants
determined with $\Lambda_{2/3}=0.5$ GeV and obtain two poles
$(M,\Gamma)$ in the non-strange channel,
\begin{equation}
(10612.0,32.2\text{) MeV},\quad(10656.9,32.3)\text{ MeV},
\end{equation}
which correspond to the $Z_c(10610)$ and $Z_c(10650)$. We also
predict two strange hidden bottom states $Z_{bs}$ and
$Z_{bs}^\prime$ near the $B_s\bar{B}^*/B^*_s\bar{B}$ and
$B_s^*\bar{B}^*$ threshold respectively,
\begin{eqnarray}
(10699.9,32.3\text{) MeV},\quad(10747.9,32.2)\text{ MeV}.
\end{eqnarray}
The resonance $Z_{bs}$ could be searched for in the $B_s\bar{B}^*$,
$B^*_s\bar{B}$, $\Upsilon K$ final states and $Z_{bs}^{\prime}$ in
the $B_s^*\bar{B}^*$, $B_s\bar{B}^*$, $B^*_s\bar{B}$, $\Upsilon K$
final states.

{\it Summary and Outlook.}---In summary, we perform the
$J/\psi\pi(K)$, $\bar{D}_{(s)}{D}^*/\bar{D}_{(s)}^*{D}$ and
$\bar{D}_{(s)}^*{D}^*$ coupled-channel calculation in the contact
interaction with the $\text{SU(3)}_F$ symmetry and HQSS. We fit the
masses and widths of $Z_{c}(3900)$ and $Z_c(4020)$ to determine the
coupling constants. We reproduce the mass and width of
$Z_{cs}(3985)$ very well as the $U/V$-spin partner of $Z_c(3900)$.
We also obtain the ratios of the partial decay widths of
$Z_{cs}(3985)$ and obtain its dominant $\bar{D}_sD^*/\bar{D}^*_s{D}$
decay modes. We introduce the $G_{U/V}$ parity to label $Z_{cs}$
states. In the $SU(3)_F$ limit, the partial decay widths to
$\bar{D}_sD^*$ and $\bar{D}^*_s{D}$ are equal. We also predict the
$Z_{cs}^{\prime}$ with the mass around 4130 MeV, which are the HQSS
partner states of the $Z_{cs}(3985)$ and $U/V$-spin partner states
of the $Z_c(4020)$. With the HQFS, we reproduce the masses and
widths of $Z_b(10610)$ and $Z_b(10650)$ and predict their $G_{U/V}$
partner states $Z_{bs}$ and $Z_{bs}^\prime$ with a mass around 10700
MeV and 10750 MeV.

\vspace{0.5cm}

\begin{acknowledgements}
We are grateful to Rui Chen for helpful discussions. This project is
supported by the National Natural Science Foundation of China under
Grant 11975033. This work is supported in part by DFG and NSFC
through funds provided to the Sino-German CRC 110 ``Symmetries and
the Emergence of Structure in QCD" .
\end{acknowledgements}

%\bibliographystyle{apsrmp4-1.bst}
%\bibliography{C:/Users/dreamway/OneDrive/Academic/Documents/paper}
\bibliography{ref}

\end{document}